\documentclass[%
reprint,
pre,
showpacs]{revtex4-1}

\usepackage{amssymb,amsmath}
\usepackage{graphicx}
\usepackage{epstopdf}
\usepackage{color}
\usepackage{url}

\begin{document}

\title{Nonlinear self-collimated sound beams in sonic crystals}

\author{El Mokhtar Hamham}
\affiliation{Electronics and Microwaves Group, Faculty of Sciences, Abdelmalek Essaadi University, Tetouan, 93000 Morocco}

\author{No\'e Jim\'enez}
\affiliation{Instituto de Investigaci\'on para la Gesti\'on Integrada de Zonas Costeras, Universitat Polit\`ecnica de Val\`encia, Paranimf 1, 46730 Grao de Gandia, Spain}

\author{Rub\'en Pic\'o}
\affiliation{Instituto de Investigaci\'on para la Gesti\'on Integrada de Zonas Costeras, Universitat Polit\`ecnica de Val\`encia, Paranimf 1, 46730 Grao de Gandia, Spain}

\author{V\'ictor S\'anchez-Morcillo}
\affiliation{Instituto de Investigaci\'on para la Gesti\'on Integrada de Zonas Costeras, Universitat Polit\`ecnica de Val\`encia, Paranimf 1, 46730 Grao de Gandia, Spain} 

\author{Llu\'is M. Garc\'ia-Raffi}
\affiliation{Instituto Universitario de Matemática Pura y Aplicada, Universitat Polit\`ecnica de Val\`encia, Cam\'{\i} de Vera s/n, 46022 Val\`encia, Spain}

\author{Kestutis Staliunas}
\affiliation{ICREA, Universitat Polit\`ecnica de Catalunya, rambla Negridi, Terrassa, Spain}

\date{\today}

\begin{abstract}
We report the propagation of high-intensity sound beams in a sonic crystal, under self-collimation or reduced-divergence conditions. The medium is a fluid with elastic quadratic nonlinearity, where the dominating nonlinear effect is harmonic generation. The conditions for the efficient generation of narrow, non-diverging beam of second harmonic are discussed.  Numerical simulations are in agreement with the analytical predictions made, based on the linear dispersion characteristics in modulated media and the nonlinear interaction in a quadratic medium under phase matching conditions.
\end{abstract}

\pacs{--}

\maketitle

\section{Introduction}

 The beams of different kind of waves, such as electromagnetic or sonic, experience diffraction and broaden as they propagate in a homogeneous medium. This spreading of energy in space diminishes the wave amplitude of the beam in the axis along propagation, unless the spreading is balanced by some focusing mechanism. Still, it is possible to prevent this fundamental wave propagation property and create non-diverging beams. Among the beam patterns without divergence the most popular are the linear Bessel beams \cite{Drunin87}, and the nonlinear solitonic or self-trapped beams \cite{Stegeman99}. Recently, another method for creating linear non-diverging beam was proposed, for waves propagating in a periodic medium. Such beams have been named self-collimated beams, and were first proposed for light beams in photonic crystals \cite{Kosaka1999}, and later extended to other type of waves. The phenomenon of self-collimation has attracted much attention, as a technique to propagate optical, acoustical or even matter wave beams at long distances without a sensible loss of amplitude. Self-collimation is highly sensitive to the frequency: the divergence of beams can be reduced or even suppressed only at particular frequencies, those presenting particular dispersion characteristics, namely flat regions in the isofrequency contours \cite{Kosaka1999}. The size of the self-collimated beam is also limited by the extension of such flat region in angular space. 
 Self-collimation of low amplitude (linear), monochromatic acoustic waves has been demonstrated in 2D \cite{Espinosa2007} and 3D \cite{Soliveres2009} sonic crystals. More recently, the simultaneous self-collimation of two beams of different frequencies was also demonstrated experimentally \cite{Soliveres2011}. These results show that the conditions for self-collimation can be achieved also for non-monochromatic beams; in particular the case of the superposition of beams of one frequency and its second harmonic was considered in \cite{Soliveres2011}. The latter results are valid in linear regime; actually in \cite{Soliveres2011} both frequency components were present in the input beam, and the corresponding beams propagated in the crystal without nonlinear interaction between them. 
 
 In the linear case, the propagation of light and sound beams obey similar equations, and similar propagation characteristics are expected. The similarities between photonic and sonic crystals are well established \cite{Miyashita05}, and have motivated many studies, where analogous effects in both systems have been investigated. The analogy, however, breaks for high amplitude waves, where nonlinear effects appear. For example, second and higher harmonic generation processes may be essentially different in optics and acoustics. One reason is the absence of intrinsic dispersion for acoustic waves propagating in homogeneous media. Nonlinear acoustical waves in nondispersive media as homogeneous fluids, eventually generate shock waves, which are not observed in optics. Also, the type and strength of nonlinearity may be different. While most common optical nonlinearities are cubic (kerr-type), in fluids and homogeneous solids, quadratic nonlinearity is dominant in acoustics. Even, nontraditional acoustic nonlinearities (power-law, hysteretic,...) are typical of some complex or microstructured acoustic media. In this sense, nonlinear effects of acoustic waves in periodic media, and in particular the self-collimation problem considered here are not a direct extension of the same effects in the optical case. Furthermore, the propagation of nonlinear acoustic beams in sonic crystals has never been addressed before.

The basic effect in nonlinear acoustics is harmonic generation \citep{Hamilton2008}. It is known that efficient harmonic generation is only possible under fulfilment of phase matching conditions. For acoustic waves in fluids, this condition is rather natural, being always fulfilled for all harmonics due to the absence of dispersion, however in optics this requires special materials and special phase matching techniques \cite{Boyd2003}. 

Acoustic harmonic generation has been studied in a variety of highly-dispersive nonlinear media, as bubbly liquids, or acoustic waveguides \cite{Hamilton2008}, and weakly dispersive media as elastic plates \cite{DeLima2003, Muller2010}, nonlinear porous-elastic media \cite{Donskoy1997} and in granular media \cite{Legland2012}. It has been proven as a useful effect in different  applications, as material characterization \cite{Hirsekorn1994,Zheng1999}, ultrasound imaging and echography, \cite{Humphrey2000}, biological tissue characterization \cite{Law1981} and other medical ultrasound applications.

The purpose of this paper is to study nonlinear propagation of high intensity sound beams in periodic media, and in particular to demonstrate the formation of nonlinear  self-collimated acoustic beams, and discuss the conditions under which this process occurs with maximal efficiency. A sonic crystal is designed, by using an iterative method, to fulfil the three conditions for optimal energy transfer between harmonics: flatness of isofrequency contour for each harmonic, phase matching and large overlap between distributions of the interacting Bloch modes. The predictions are checked by FDTD simulations of the nonlinear problem, that demonstrate the efficient generation of fundamental and second harmonic acoustic narrow beams. 

\section{Nonlinear sound beam propagation model}

Several models can be used to describe nonlinear sound wave propagation though a fluid medium, with different levels of accuracy. An accurate description, when thermal and viscous effects are negligible, follows from the conservation laws of mass and momentum, can be written, respectively, in a Eulerian form \cite{Hamilton2008}: 
\begin{eqnarray}
	\frac{{\partial \rho }}{{\partial t}} =  - \nabla  \cdot \left( {\rho {\bf{v}}} \right),\label{eq_mass}\\
	\rho \left( {\frac{{\partial {\bf{v}}}}{{\partial t}} + {\bf{v}} \cdot \nabla {\bf{v}}} \right) =  - \nabla p,\label{eq_momentum}
\end{eqnarray}
where $\textbf{v}$ is the particle velocity vector, $p$ is the acoustic pressure, $\rho$ is the total density field that can be expressed as $\rho = \rho_0 + \rho'$, where $\rho _0$ the ambient fluid density and $\rho'$ is the acoustic density. The system is closed by the equation of state of the fluid, that under our assumptions is a pressure-density relation, $p=p(\rho)$. A commonly used expression is obtained after Taylor expansion, keeping nonlinear terms up to second order. Then 
\begin{equation}\label{eq_state}
p = c_0^2 \rho + \frac{c_0^2}{\rho_0}\frac{B}{2A}\rho^{2},
\end{equation}
where $B/A$ is the nonlinear parameter of the medium (which is known for most of materials, see e.g. \citep{Naugolnykh1998}) and $c_0$ the sound speed in the medium.
Note that Eqs. (\ref{eq_mass}) and (\ref{eq_momentum}) also contain nonlinearities related to (1) mass and momentum advection , or (2) geometrical nonlinearities. However, for the second harmonic generation they are of minor importance compared with the nonlinear terms in the equation of state, Eq. (\ref{eq_state}). 

 The above formulation of nonlinear propagation problem remains valid when the propagating medium is inhomogeneous, including the case of sonic crystals where inhomogeneity is periodically distributed in space. In such case, the medium parameters $c_0$ and $\rho _0$ are space-dependent, represented by periodic functions. To our knowledge, the propagation of acoustic beams in periodic media has been only studied in the linear regime, and the corresponding nonlinear problem is addressed here for the first time.
 
\begin{figure}[t]
	\includegraphics[width=8cm]{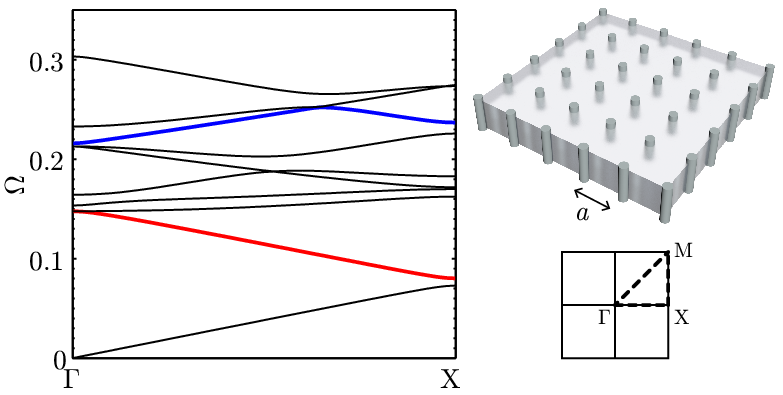}\\
	\includegraphics[width=4.1cm]{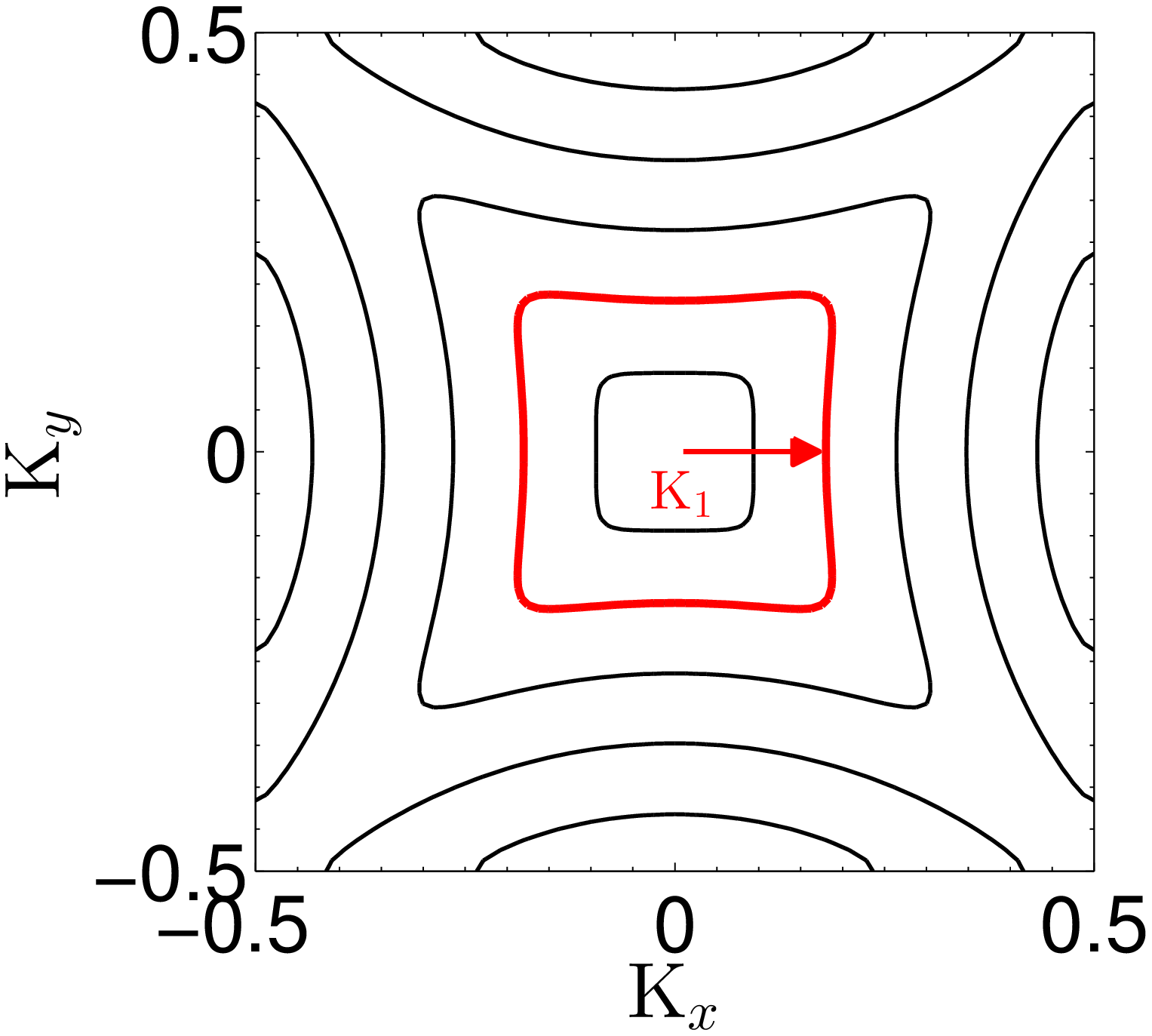}
	\includegraphics[width=4.1cm]{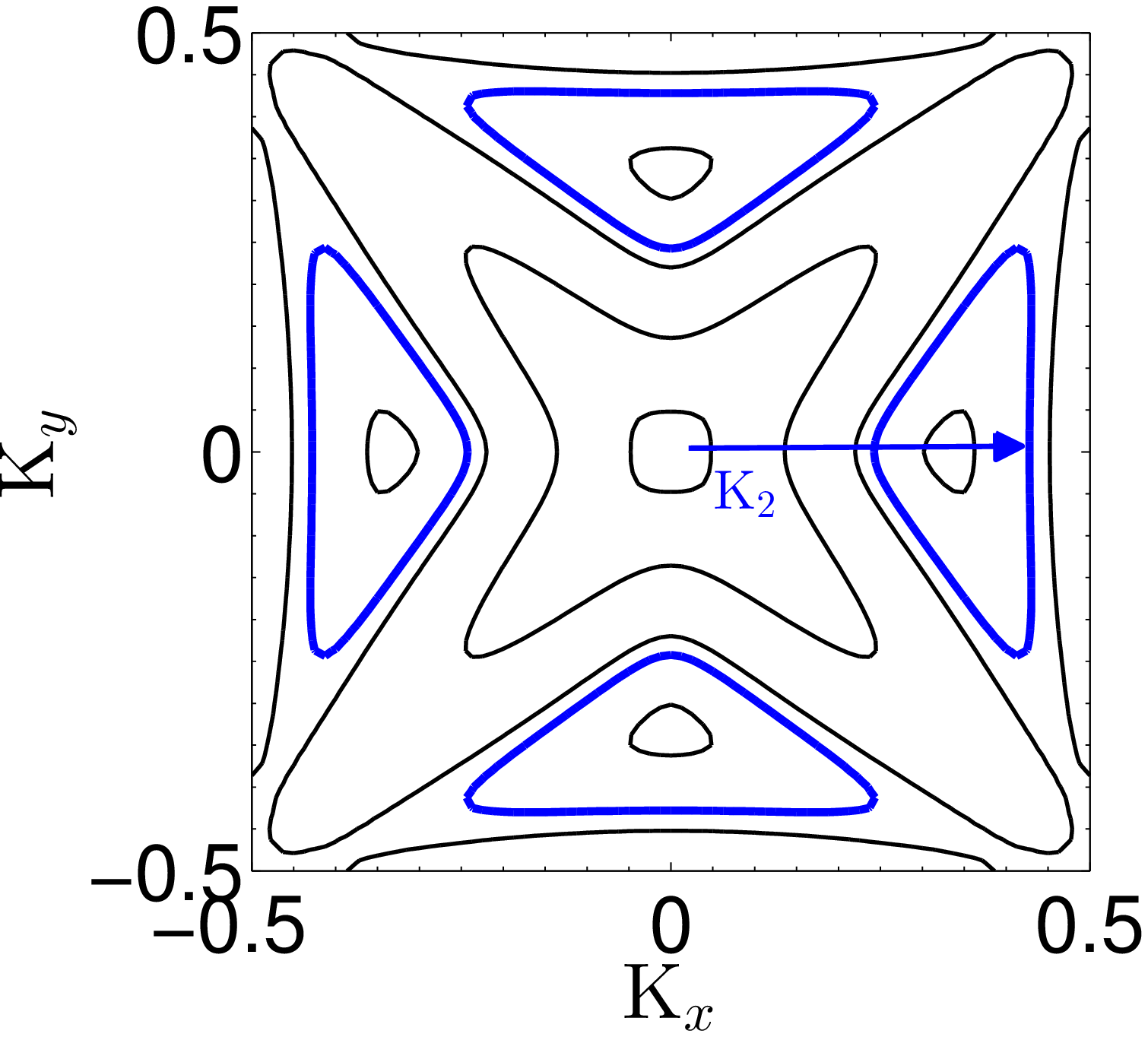}

	\caption{Top: Band structure (left) of a square lattice of rigid cylinders with $r = 0.11 a$, where $a$ is the lattice constant, immersed in water (right). Red and blue lines mark the bands (2nd and 8th) for which simultaneous self-collimation for both fundamental and second harmonic searched, along $\Gamma-\mathrm{X}$ direction. Bottom: Isofrequency contours for the 2nd (left) and 8th bands (right).Points denote the wavevectors for both waves, lying on flat segments respectively}
	\label{FigDiagram}
\end{figure}

\section{Self-collimation of intense acoustic beams}

We consider a narrow, intense acoustic beam incident on a 2D sonic crystal made of cylindrical scatterers of radius $r$ embedded in a fluid, arranged in a square-lattice with lattice constant $a$. The corresponding filling fractio is $f=\pi (r/a)^2$. The beam width is roughly $6$ lattice periods. For the sake of simplicity the scatterers are considered perfectly rigid (the sound field is totally reflected from the wall of scatterer). Assuming water as a host fluid, the material parameters are $\rho _0 = 1000$ kg/m$^3$, $c_0 = 1490$ m/s.

The special conditions required for a sound beam to propagate without diffraction are presented in this section. The problem of self-collimation has been already discussed for linear, monofrequency \cite{Perez2007,Espinosa2007} and bi-frequency \cite{Soliveres2011} beams. Since a nonlinear beam is composed by a fundamental frequency component and its high frequency harmonics, self-collimation of the nonlinear beam requires self-collimation of its constituent frequency components. We remind that in self-collimation regime the sonic beam does not spread diffractively because Bloch wave vectors lying on the flat segment of the spatial dispersion curve have equal longitudinal components and thus do not dephase mutually in propagation. In general, this flatness of the dispersion curve appears at a particular frequency, but as shown in \cite{Soliveres2011} it can be also obtained for a wave and its second harmonic regarded they propagate in different propagation bands. 

To illustrate this case, we show in Fig. \ref{FigDiagram} the dispersion diagram of the sonic crystal for small-amplitude excitations, obtained using the Plane Wave Expansion (PWE) method on a linearized version of Eqs. (\ref{eq_mass})-(\ref{eq_state}). The conventional form of the band diagram is represented on the trajectory along the principal directions of the crystal, $\Gamma-\mathrm{X}-\mathrm{M}-\Gamma$, which are the boundary of the irreducible Brillouin zone (BZ). Figure~\ref{FigDiagram} (top) shows the dispersion diagram for $\Gamma-\mathrm{X}$ direction. The red line in Fig.~\ref{FigDiagram} denotes the 2nd propagation band. The fundamental (driving) field lies on this band. Its frequency $\Omega$ is chosen such that the corresponding isofrequency contour contains flat regions. The blue line in Fig.~\ref{FigDiagram} denotes the 8th propagation band. The second harmonic frequency, $2\Omega$, lies in this band for the particular crystal parameters considered.  
As shown in \cite{Soliveres2011}, for a given crystal, it is possible to choose the fundamental frequency such that the isofrequency contours for both frequencies present flat regions. 
In our particular crystal, this happens when $\Omega$ = 0.125 [these are dimensionless frequencies, related to physical frequencies $\omega$ as  $\omega=\Omega (2 \pi c_0/a)]$. Similarly, a normalized Bloch wavevector is defined as $\mathrm{K}=\mathrm{k}_x(a/\pi)$

Such condition is necessary to achieve self-collimated propagation of the nonlinear beam.     
In order to obtain an efficient generation of the second harmonic, together with simultaneous self-collimation for both waves two additional geometric conditions must be fulfilled, related to the wavenumber and spatial shape of the interacting beams. 
These conditions have been discussed for photonic crystals \cite{Nistor2008, Nistor2010}. 

\subsection{Phase matching conditions}

 It is well known from nonlinear optics \cite {Boyd2003} that a proper phase relationship between the fundamental and second harmonic waves must be satisfied for an efficient nonlinear frequency conversion along the propagation direction. In a dispersive medium, the wavenumbers of first harmonics do not combine to result precisely in wavevector of second harmonics, and a phase mismatch $\Delta \mathrm{K}=2\mathrm{K}(\Omega)-\mathrm{K}(2\Omega)$ occurs. As a consequence, the second harmonic field is limited in amplitude: it does not grow linearly but oscillates in propagation, with a characteristic period given by the coherence length $l_c=\frac{\pi}{\Delta \mathrm{k}}=\frac{a}{\Delta \mathrm{K}}$ \citep{Boyd2003}.  
 
 The conversion efficiency into second harmonics generally is smaller in optics than in acoustics, because of the inherent material dispersion for light waves (absent for sound waves in fluids), that causes the fundamental and second harmonic waves to travel along the crystal with different phase velocities. Thus, the presence of the scatterers is the only important source of dispersion in the acoustic case.

\begin{figure}[tb]
	\centering
	\includegraphics [width=8cm]{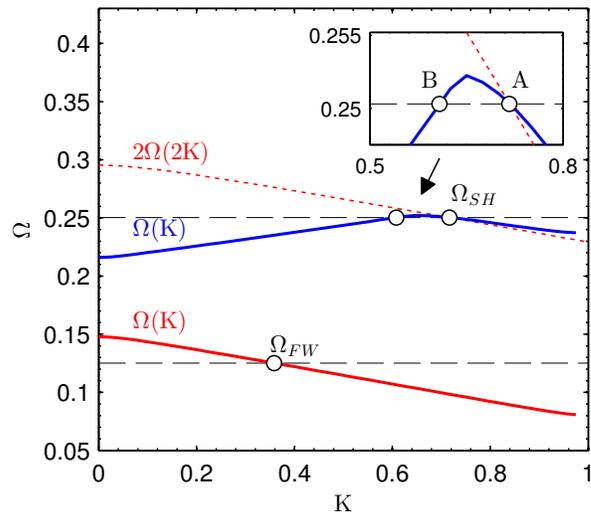}
	\caption{Dispersion curves involved in simultaneous self-collimation for the fundamental (red line) at the 2nd band and second harmonic (blue line) at the 8th band. The "doubled" dispersion curve is represented (dashed red line) to identify phase matching of harmonics. The intersection denotes the frequency presenting phase matching. A closest view (inset) shows that for the self-collimated second harmonic, two solutions (modes A and B, with distinct $\mathrm{K}$) are found, phase and non-phase matched, respectively}
	\label{SHmatch}
\end{figure}

Phase matching corresponds to $\Delta \mathrm{K} = 0$. Figure ~\ref{SHmatch} shows that it can be actually achieved for the pair of frequencies where self-collimation occurs, as follows from the previous analysis. There, we represent the dispersion branches involved in self-collimation, along $\Gamma-\mathrm{X}$ direction, as in Fig.~\ref{FigDiagram}. Fundamental and second harmonic modes correspond to the crossings of the dotted horizontal lines with the corresponding dispersion branches. For a given fundamental frequency $\Omega$, in order to check the fulfillment of the phase matching condition, the curve corresponding to the ''double'' of the 2nd band dispersion curve, $2\Omega(2\mathrm{K})$, has been represented in Fig.~\ref{SHmatch} as a red dashed line. Phase matching  is satisfied at the intersection between this curve with the corresponding curve at the 8th band. This corresponds to the mode labeled A in Fig.~\ref{SHmatch}, which is phase-matched with the fundamental mode.

Note that due to the concavity of the 8th band, at frequency 2$\Omega$ a second mode labelled with B in Figure.~\ref{SHmatch} can be also excited. This solution presents a large phase mismatch with the fundamental mode, and its contribution to the second harmonic field is negligible. 

The simultaneous fulfillment of both conditions is obtained by an iterative procedure, which implies a re-design of the crystal parameters. The procedure is as follows: we start from a pair of frequencies $(\Omega,2\Omega)$ showing self-collimation for a given crystal parameters. Around this doublet, we seek the closest pair of frequencies $\Omega'=\Omega+\delta \Omega$ and 2$\Omega'$ showing phase-matching. Then the isofrequency curves (Fig.~\ref{FigDiagram}) are again calculated in order to evaluate the deviation of flatness in the isofrequency contours. The sonic crystal parameters are then modified, e.g. by a slight variation of the filling fraction, in order to tune the dispersion relations to get again self-collimation conditions. The process is repeated again until both conditions (flatness and phase matching) are simultaneously satisfied. We note that, despite this is out of the scope of this paper, optimization techniques as genetic algorithms can be applied here in order to find an optimal structure.

\subsection{Nonlinear coupling of Bloch modes}

\label{sec:Nonlinearcoupling}
\begin{figure}[t]
	\centering
	\includegraphics[width=4.1cm]{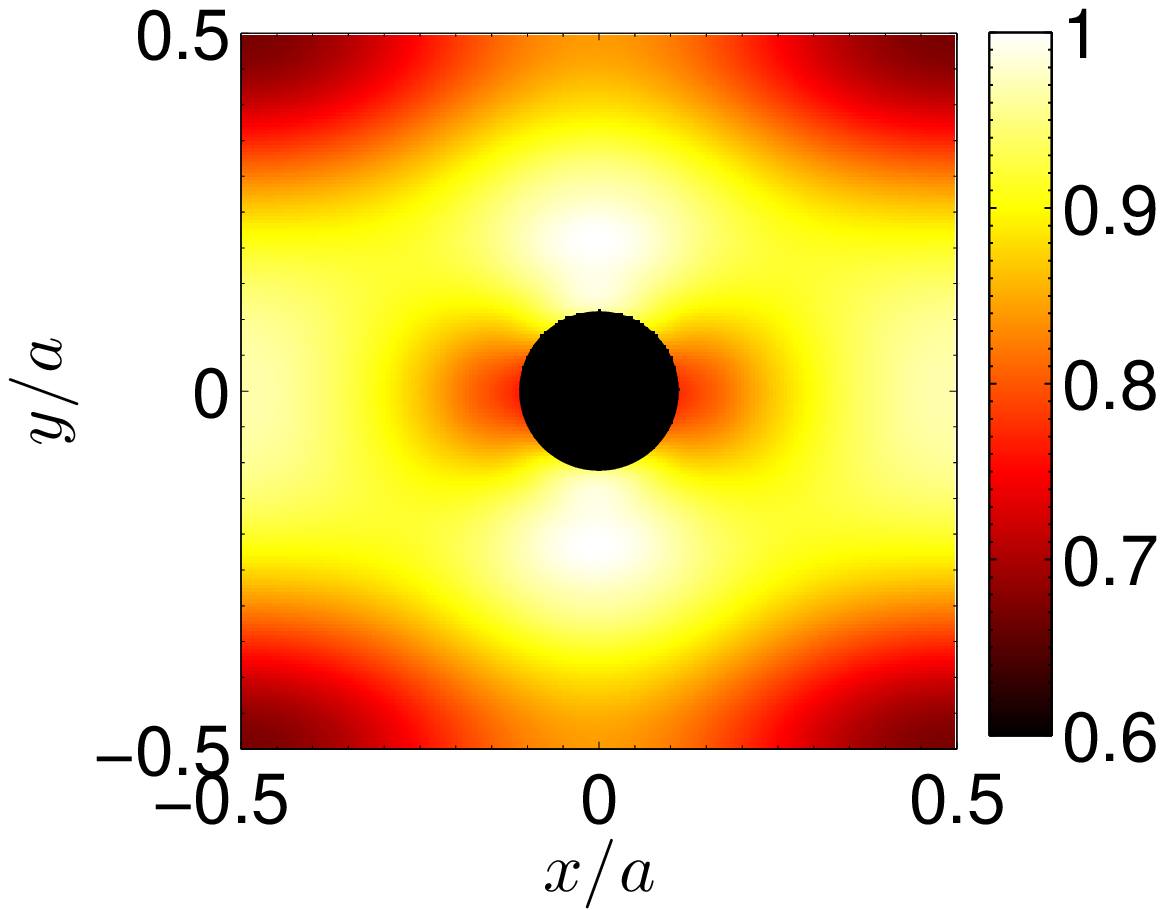}
	\includegraphics[width=4.1cm]{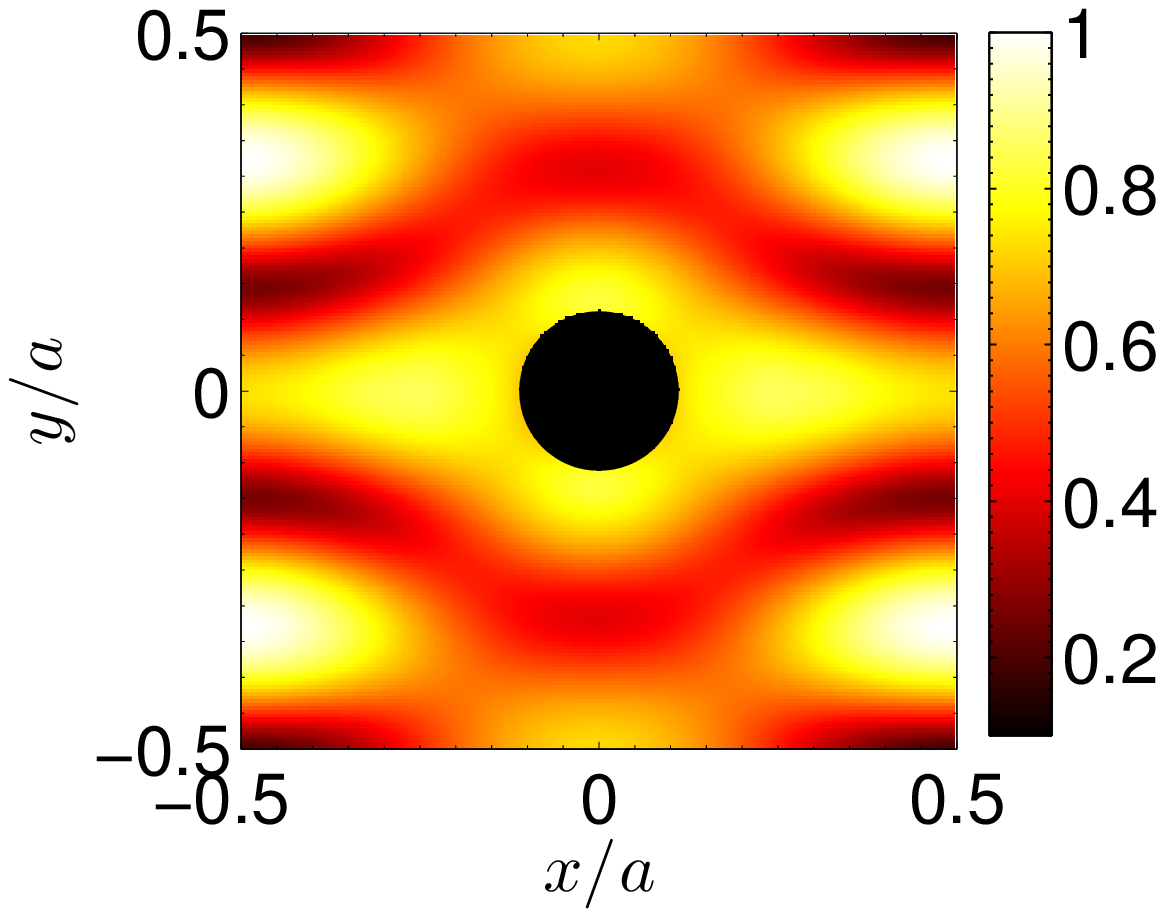}
	\caption{Spatial distribution of the pressure field for the Bloch modes of the findamental (left) and the second harmonic (right) waves. The coupling coefficient is estimated to be  $\kappa = 0.85$.}
	\label{BlochMode}
\end{figure}

Efficient energy transfer between harmonics requires also a strong mode coupling, which depends on the spatial overlapping between the two interacting waves. For plane waves in a homogeneous medium a perfect spatial nonlinear coupling between first and second harmonic is assured, since mode overlapping is maximal. The propagation eigenmodes in a periodic medium are Bloch waves, whose amplitudes are spatially modulated and does not necessarily overlap. If two modes do not overlap in space, the energy transfer is less efficient even if they are phase matched. The amount of energy transfer can be estimated evaluating the spatial overlap between the envelopes of the corresponding Bloch modes. Let $B_{1}$ and $B_{2}$ be the spatial envelopes of the Bloch modes of the fundamental and and second harmonic waves, respectively. The nonlinear coupling coefficient is calculated as the cross-correlation between the functions $B_{1}^{2}$ and $B_{2}$ normalized in such a way that unity would correspond to the perfect matching of the modes \cite{Nistor2008}. We define the coupling coefficient as
\begin{equation}
\kappa = \frac{| \int_{M}  B_{1}^{2}\,B_{2}^{*}  \mathrm{d}r |} { \sqrt{ \int_{C} \, | B_{1}^{4}|\,\mathrm{d}r \, \int_{C} \, | B_{2}|^{2}\,\mathrm{d}r }}
\end{equation}
where the upper integral is calculated in the nonlinear medium from one unit cell, while the lower integrals are taken over the entire unit cell. 
To calculate $B_{1}$ and $B_{2}$ we solve the eigenvalue problem for the pressure field by means of the PWE method, which converts the differential equation to an infinite matrix eigenvalue problem that can be truncated and solved numerically. For that, we follow the same procedure as in \cite{Perez2007} however the problem is solved inversely, i.e. for a given frequency, the corresponding wave vectors, satisfying the phase matching condition, are obtained. Then $B_{1}$ and $B_{2}$ are obtained as the eigenvectors corresponding to fundamental and second harmonic frequencies, respectively. In Fig.~\ref{BlochMode} we plot the spatial distributions of $B_{1}$ and $B_{2}$, respectively, for the selected final design where a coupling coefficient of $\kappa = 0.85$ is obtained. This value is of the same order as the coupling in homogeneous media, $\kappa = 1$, and therefore sufficiently large for an efficient harmonic generation.

\subsection{Numerical simulation}

\begin{figure}[t]
	\centering
	\includegraphics[width=4cm]{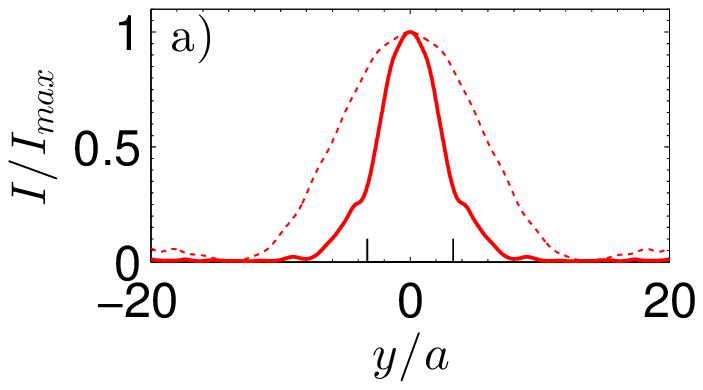}
	\includegraphics[width=4cm]{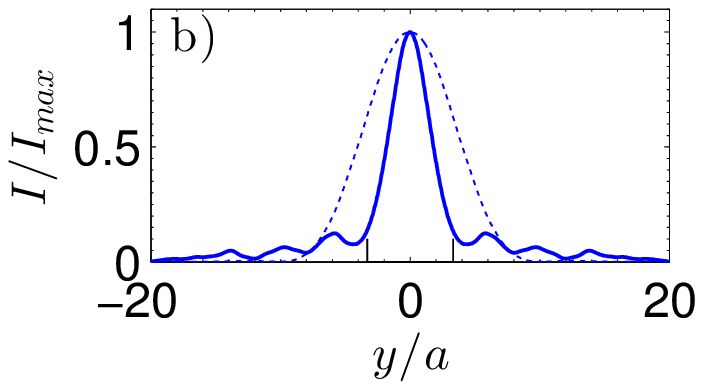}\\
	\includegraphics[width=4cm]{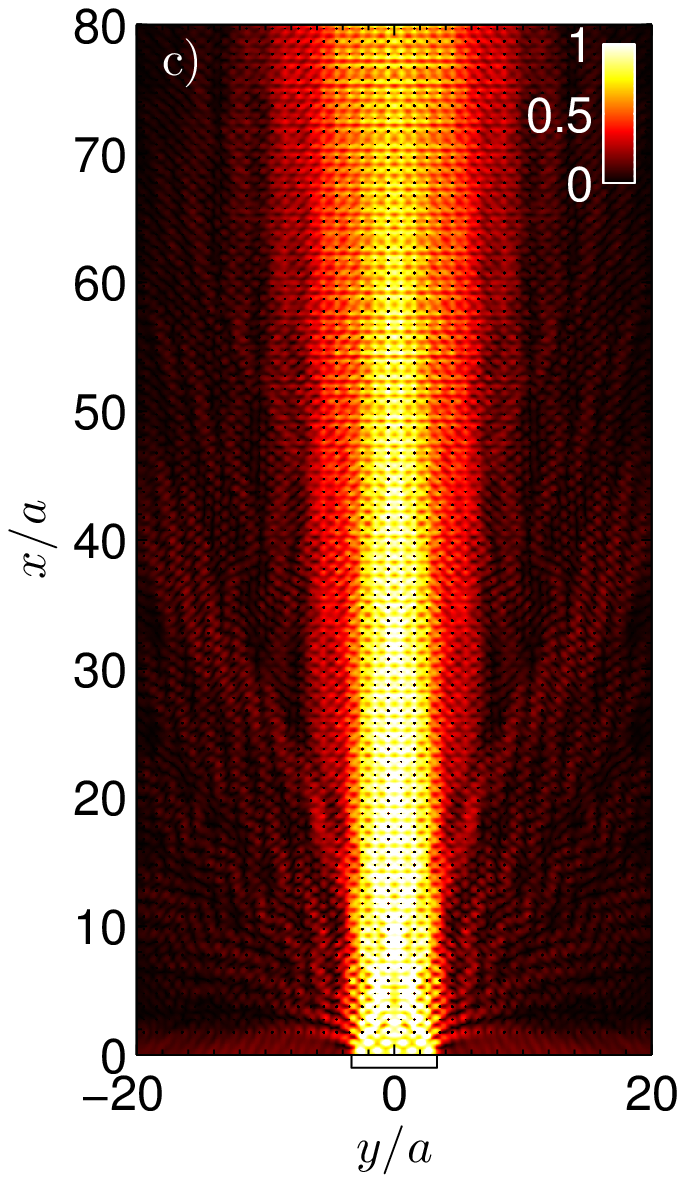}
	\includegraphics[width=4cm]{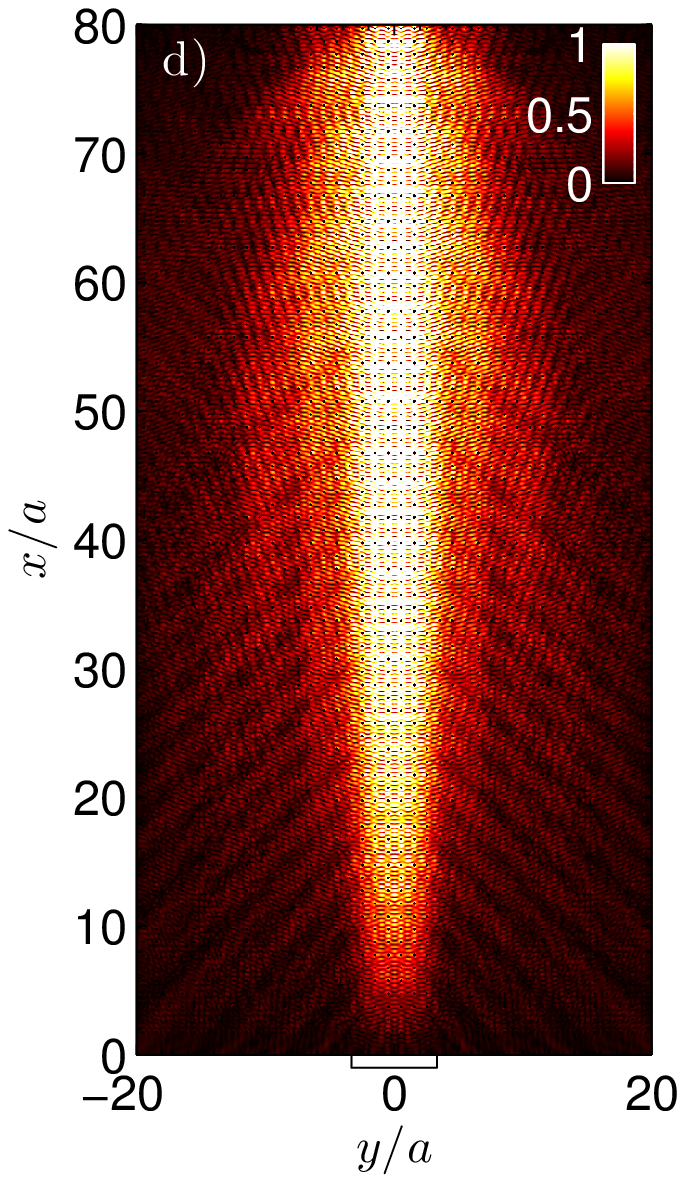}\\
	\includegraphics[width=4cm]{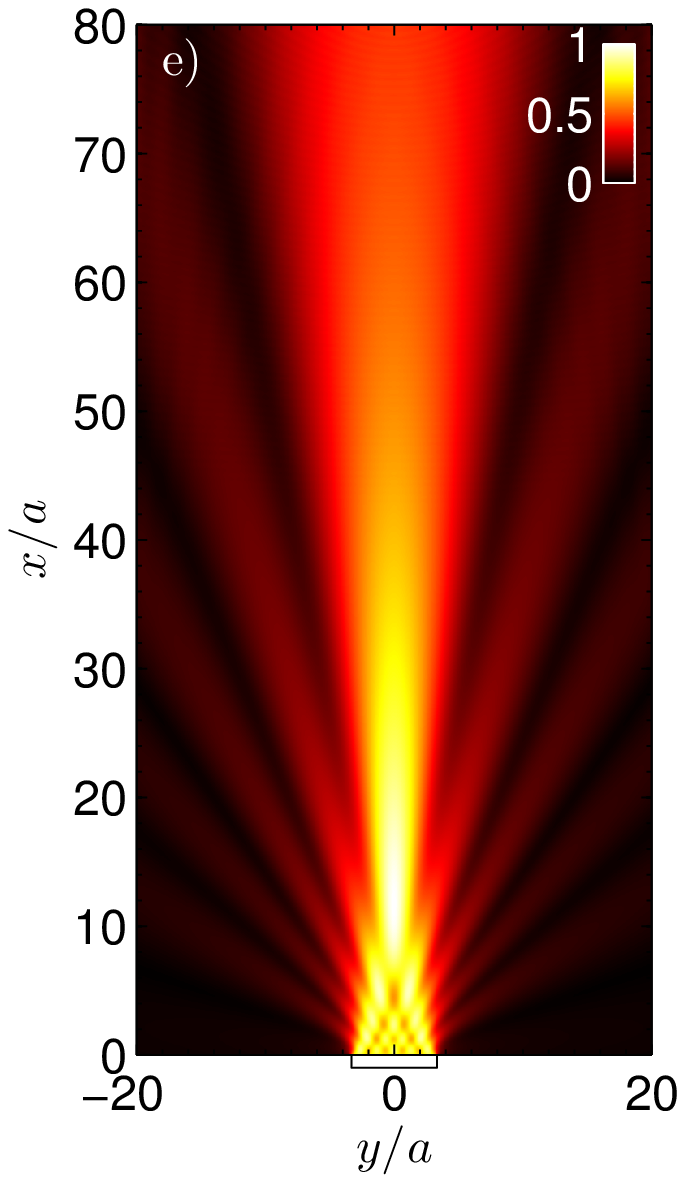}
	\includegraphics[width=4cm]{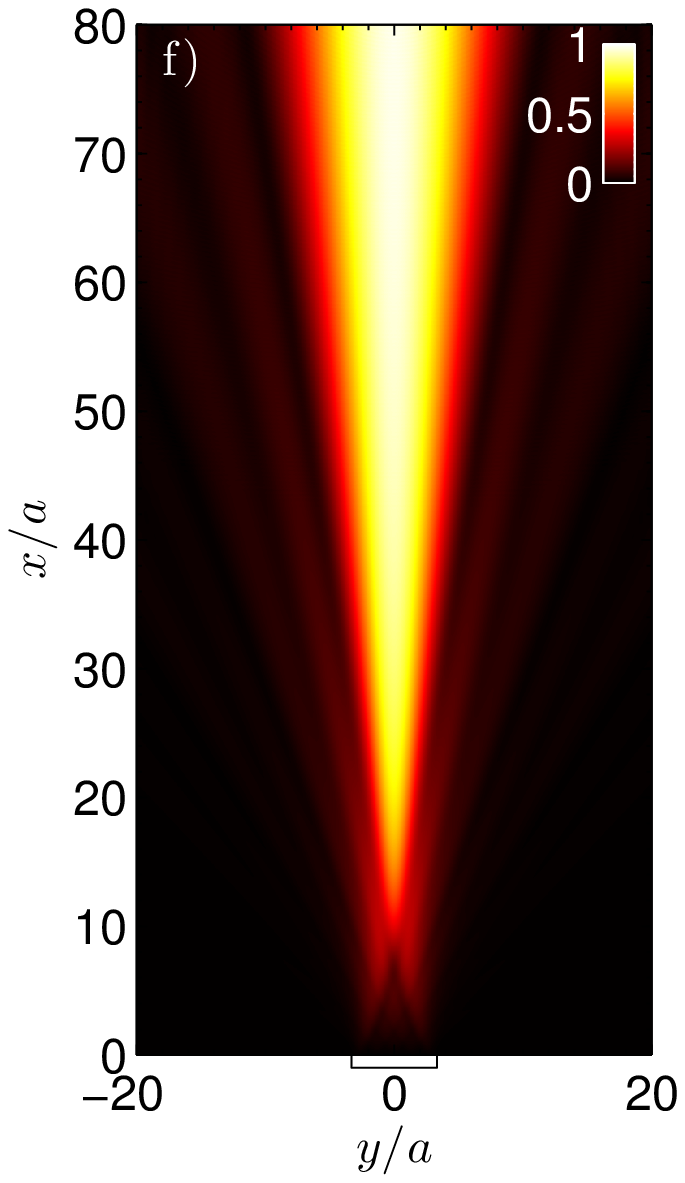}
	\caption{Pressure distributions obtained by FDTD simulations. Normalized intensity cross section of fundamental (a) and second harmonic (b) at $x=80a$ propagating in the crystal (continuous line) and in a homogeneous (water) medium (dashed lines). Beam spatial distribution for simultaneously self-collimated harmonics in the sonic crystal (c-d) and in homogeneous fluid (e-f). Pressures are normalized to the maximum pressure.}
	\label{figFDTD}
\end{figure}

A full-wave nonlinear simulation was performed, using the FDTD method, to validate the efficiency of the second harmonic generation in the proposed structure. The crystal parameters, obtained after the iterative procedure describe above, are as in Figs. (\ref{FigDiagram}) and (\ref{SHmatch}). The source is a plane piston with a width of $6a$, located near the crystal, and radiating a harmonic wave with normalized frequency $\Omega$ = 0.125. In order to minimize numerical dispersion a computational grid with $N_{\lambda}=45$ elements per wavelength was used, and a Courant-Friedrich-Levy number of $S=0.95$. In Fig.~\ref{figFDTD} we present the numerically obtained spatial distributions of the fundamental and its nonlinearly generated second harmonic. As predicted, both beams are nearly collimated. For comparison, the beam spatial distribution calculated for an homogeneous material (removing the crystal) are represented in Fig.~\ref{figFDTD}~(e-f), where the diffractive broadening of the beams is visible. Transversal intensity distributions are shown in Fig.~\ref{figFDTD}~(a-b) for a distance $80a$, where beam widths are compared with the reference beams in the homogeneous medium, broadened by diffraction. 

\begin{figure}[t]
	\includegraphics[width=8cm]{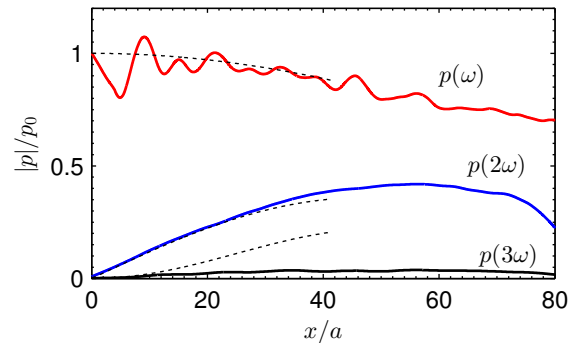}
	\centering
	\caption{Normalized field amplitude along the acoustical axis $y=0$, for the fundamental, second and third harmonic beams. The dashed lines represent the analytical solutions for harmonic evolution on a plane wave propagating in a nondispersive medium.}
	\label{FWSHout}
\end{figure}

The pressure amplitudes along the beam axis are shown in Fig.~\ref{FWSHout} for each harmonic. Here, the analytic solution for a nonlinear plane wave propagating in a homogeneous (nondispersive) medium is plotted for reference (dashed lines) \cite{Naugolnykh1998}, given by $p_n/p_0=2 J_n(n\sigma)/n\sigma$, where $J_n$ is the Bessel funcion of order $n$, $\sigma=(\varepsilon \omega p_0/\rho_0 c_0^3) x$ is coordinate normalized to the shock formation distance, $p_0$ is the pressure at the source and $\beta=1+ B/2A$ the nonlinearity parameter. Such analytical solution is valid in the pre-shock region $\sigma<1$. The growth rate of the self-collimated second harmonic beam propagating in the crystal matches well the growth rate of a plane wave in a homogeneous medium in such preshock region, which is a consequence of the weak divergence of the beam and the high degree of phase matching. Also, the second harmonic field can reach even higher amplitudes than those corresponding to nondispersive media (where harmonics decay beyond the shock formation distance). The latter effect can be understood in terms of phase mismatch of higher harmonics: in homogeneous media all harmonics are phase matched while in the crystal only the second harmonic is phase matched. In Fig. (\ref{FWSHout}) the third harmonic is also plotted, where its small contribution to the beam is evident. The phase mismatch in third and higher harmonics decrease the energy flow into these components. Finally, the amplitude in Fig. (\ref{FWSHout}) decays because non-perfect conditions for self-collimations, that makes the beam to start diverging after a long distance, or non-perfect phase matching, which results in a beating period with long coherence length.

\section{Conclusions and Remarks}

We have demonstrated the possibility of efficient second harmonic generation of sound in a sonic crystal, by means of the formation of narrow, weakly diverging nonlinear acoustic beams. Three conditions must be simultaneously present for a efficient second harmonic generation, which are: 1) simultaneous self-collimation, 2) phase mathing and 3) high spatial coupling of interacting harmonics. The use of simultaneous self-collimation regime limits the diffraction of both harmonic beams, maintaining the amplitude at the axis and therefore the nonlinear interaction. Under ideal conditions (no divergence and losses), the decrease of the first harmonic beam is mainly attributed to the energy transfer to second and higher harmonics. The sonic crystal parameters can be chosen to fulfill phase matching with the second harmonic, maximizing second harmonic generation due to synchronous cumulative interaction. Finally, the spatial coupling (overlapping) between interacting modes is also analyzed by calculating a nonlinear coupling coefficient. It is shown that its value ($\kappa=0.85$ for the case studied) is not far from the ideal case, revealing a strong spatial overlap between both Bloch modes that leads to high energy transfer.

 The study show that linear dispersion characteristics (band structures, isofrequency contours) can be used to predict the behaviour of nonlinear beams propagating in periodic media. This opens the possibility of extending the study of nonlinear sound beam propagation in sonic crystals to other cases of interest. For example, crystals with higher filling factors present full frequency bandgaps, that may be used to filter out the propagation of selected higher harmonics. In this sense, sonic crystals can be a way to control the spectrum of intense acoustic waves, using the strong dispersion properties introduced by the periodicity.

\bibliographystyle{apsrev}

\end{document}